\documentclass[a4,12pt]{article}
\usepackage{amssymb,amsmath, amsthm}
\usepackage{verbatim}
\usepackage{color}
\newcommand{\be}{\begin{eqnarray}}
\newcommand{\ee}{\end{eqnarray}}
\newcommand{\del}{\partial}
\newcommand{\hmu}{\hat{\mu}}
\newcommand{\Tr}{\mathrm{Tr}}

\newcommand{\vev}[1]{\langle{#1}\rangle}
\makeatletter
\newcommand{\fslash}[2][0mu]{%
    \mathchoice
     {\fsl@sh\displaystyle{#1}{#2}}%
     {\fsl@sh\textstyle{#1}{#2}}%
     {\fsl@sh\scriptstyle{#1}{#2}}%
      {\fsl@sh\scriptscriptstyle{#1}{#2}}}
    \newcommand{\fsl@sh}[3]{%
    \m@th\ooalign{$\hfil#1\mkern#2/\hfil$\crcr$#1#3$}}
\def\lsim{\raise0.3ex\hbox{$<$\kern-0.75em\raise-1.1ex\hbox{$\sim$}}}
\def\gsim{\raise0.3ex\hbox{$>$\kern-0.75em\raise-1.1ex\hbox{$\sim$}}}
\makeatother
\usepackage{epsfig}
\usepackage{graphics}

\begin{document}
\hfill BI-TP 2011/07
\begin{center}
{\bf\Large  
Fluctuations as probe of the QCD phase \\
transition and  freeze-out in heavy ion \\[2mm]
collisions at LHC and RHIC}

\vspace*{0.5cm}
{\large
B. Friman$^1$, F. Karsch$^{2,3}$, K. Redlich$^{4,5}$ 
and V. Skokov$^1$} \\[4mm]
\hskip -0.5cm \small $^1$ GSI Helmholtzzentrum f\"{u}r
  Schwerionenforschung , D-64291 Darmstadt, Germany,\\
\hskip -0.5cm \small $^2$ Fakult\"at f\"ur Physik, Universit\"at Bielefeld,
D-33501 Bielefeld, Germany\\
\hskip -0.5cm \small  $^3$ Physics Department, Brookhaven National Laboratory, Upton, NY 11973, USA \\
\hskip -0.5cm   \small     $^4$ Institute of Theoretical Physics University of Wroclaw,
PL-50204 Wroclaw, Poland\\
\hskip -0.5cm $^5$\small ExtreMe Matter Institute EMMI, GSI, D-64291 Darmstadt, Germany
\end{center}

\vspace*{0.4cm}
\centerline{Abstract} {\small
We discuss the relevance of higher order cumulants of net baryon number
fluctuations for the analysis of freeze-out and critical conditions
in heavy ion
collisions at LHC and RHIC. Using properties of $O(4)$
scaling functions, we discuss the generic structure of these higher cumulants
at vanishing baryon chemical potential and apply chiral model calculations
to explore their properties at non-zero baryon chemical potential.
We show that the ratios of the sixth to second and eighth to second
order cumulants of the net baryon number fluctuations change rapidly in the
transition region of the QCD phase diagram. Already at vanishing
baryon chemical potential they deviate considerably from the predictions of
the hadron resonance gas model which reproduce 
the second to fourth order cumulants of the net proton number fluctuations
at RHIC. We point out that the sixth order cumulants of baryon number and
electric charge fluctuations remain negative at the chiral
transition temperature. Thus, they offer the possibility to probe the proximity 
of the chemical freeze-out to the crossover line.
}
\newpage
\section{Introduction}

Strongly interacting matter at high temperature or large net baryon
number density is expected to undergo a rapid transition from a phase
with hadrons as dominant degrees of freedom to a phase where
partonic degrees of freedom prevail. At vanishing baryon chemical
potential ($\mu_B=0$), this transition is a true second order phase
transition only in the limit of vanishing light quark masses. For
$\mu_B > 0$, however, a second order phase transition point, the so-called
chiral critical point, may
exist also for physical values of the quark masses. A large experimental
as well as theoretical effort is put into the exploration of the QCD phase
transitions and the development of appropriate tools and observables
that can provide a univocal signal for the existence of the phase transitions
and their universal properties. The analysis of fluctuations of various
physical observables \cite{koch}, in particular of  the net baryon
number, may serve this purpose \cite{ste,raj}. Theoretically, the
properties of the corresponding susceptibilities are well understood at
high temperature  ($T$) and small values of the baryon chemical potential
($\mu_B$). In this regime, they are suitable observables for localizing the
phase boundary in the $\mu_B$-$T$ plane.

Critical behavior is signaled by long range correlations and
increased fluctuations, owing to the appearance
of massless modes at a second order phase transition.
Fluctuations of baryon number and electric charge
have been shown to be sensitive indicators for such critical behavior
\cite{lattice}.
In the exploration of the QCD phase diagram at non-zero temperature
and baryon chemical potential, higher order cumulants of baryon number
fluctuations play a particularly important role. They diverge on
the chiral phase transition line $T_c(\mu_B,m_q\equiv 0)$ as well as at
the elusive chiral critical point \cite{Stephanov}.

In heavy ion experiments, a lot of information has been collected on
particle yields in a wide range of beam energies \cite{data}. The 
particle multiplicities are well described in a thermal model using
the partition function of a hadron resonance gas (HRG) \cite{HRG}.
Ratios of particle yields at a given beam energy can be
characterized by a few thermal parameters, e.g. temperature and chemical
potentials for baryon number, electric charge and strangeness.
These parameters define the freeze-out conditions, {\it i.e.} the thermal
parameters corresponding to the last interaction of the hadrons 
participating in the collective expansion and cooling
of the hot and dense matter formed in a heavy ion collision. Data obtained at small values
of the baryon chemical potential suggest that the freeze-out curve $T_f(\mu_B)$ 
is close to the expected QCD phase boundary. In particular, at $\mu_B=0$
the chemical freeze-out seems to occur at or very  near the QCD transition 
region for a physical quark mass spectrum \cite{BraunMunzinger}.
At larger values of
$\mu_B/T$, however, there is a discrepancy between the slope of the
freeze-out curve and current lattice QCD results on the curvature of
the chiral phase transition line \cite{Mukherjee}.

The HRG model, which is based on the observed hadron spectrum,
does not exhibit critical behavior nor does it
reflect the sudden change of degrees of freedom in the
transition to the partonic phase of QCD. In the chiral limit, close to the
phase transition line $T_c(\mu_B,m_q\equiv 0)$ fluctuations of e.g. the
net baryon number density are expected to reflect the universal properties \cite{Allton}
of the 3-dimensional, $O(4)$ symmetric spin model \cite{Engels}.
The $O(4)$ scaling relations for cumulants of net baryon number fluctuations
differ significantly from predictions based on the HRG model.
Lattice calculations of cumulants of baryon number and electric charge
fluctuations, performed in the transition region at vanishing baryon chemical potential
and non-zero quark mass, do
indeed show that these cumulants differ qualitatively from those of the HRG model and reflect the
basic feature expected close to the chiral phase transition \cite{lattice}.
This suggests that at physical values of the light and strange quark masses
these cumulants are sensitive to the critical dynamics in the chiral limit
and may be employed to characterize also the
crossover transition in strongly interacting matter.

Thus, also at vanishing baryon chemical potential, {\it i.e.}
under the conditions approximately realized in the high energy runs at RHIC or LHC, 
the question arises to what extent a refined analysis of the freeze-out
conditions can establish the existence of the chiral phase transition.
At $\mu_B/T \simeq 0$, net quark number fluctuations and their higher cumulants
can be computed within the framework of lattice QCD~\cite{lattice, Gavai:2008zr}. Such calculations will
eventually provide a complete theoretical characterization of the
thermal conditions in the crossover region. This may then be used to unravel  
the relation of the freeze-out conditions at RHIC and LHC energies to the 
pseudo-critical line in the QCD phase diagram, provided the system 
remains close to thermal equilibrium during freeze-out. 
At present, however, lattice calculations provide only limited
information on cumulants up to eighth  order. 
In particular, controlled predictions on their properties in the continuum limit are
still lacking; their characteristic features are obtained on a qualitative
level, but quantitative results are not yet available.
Viable alternatives for discussing qualitative
features of the net baryon number fluctuations is offered by O(4) scaling theory and by
chiral models, like e.g. the Nambu-Jona-Lasinio  (NJL) model. In particular, the effective models have
the advantage
that they can be extended to $\mu_B > 0$ with minimal effort. On the other hand,
a clear disadvantage of NJL-type chiral models is that
they do not account for the potentially
large contribution from resonances in the hadronic phase.

In this paper we will  discuss the robust features of cumulants
of net baryon number fluctuations that can be extracted from
considerations based on $O(4)$ universality, on existing lattice
calculations and on model calculations. In the next section we
discuss higher order cumulants at vanishing baryon chemical potential
making use mainly of $O(4)$ universality. In Section 3 we
extend these considerations to $\mu_B/T > 0$ using results from
model calculations. In Section 4 we summarize the relevance of our
findings for experimental studies of baryon number fluctuations at LHC
and RHIC and in Section 5 we give our conclusions.

\section{Charge fluctuations at \boldmath $\mu_B=0$}

\subsection{\boldmath $O(4)$ scaling functions and net baryon number 
fluctuations}

Close to the chiral limit and  at temperatures near the chiral phase
transition temperature $T_c$, higher order derivatives of the free
energy density ($f$) with respect to temperature or chemical potential
are increasingly sensitive to the non-analytic (singular) part
($f_s$). We may represent the free energy density in terms of the
singular and regular contributions
\begin{equation}
f(T,\mu_q,m_q) = f_s(T,\mu_q,m_q) + f_r(T,\mu_q,m_q) \; .
\label{freeenergy}
\end{equation}
In addition to the dependence on temperature $T$, we also introduce here an explicit
dependence on the light quark chemical potential, $\mu_q=\mu_B/3$,
and the (degenerate) light quark masses $m_q\equiv m_u = m_d$. For
simplicity we do not take the chemical potentials for electric charge and strangeness into account, nor do we introduce an explicit dependence on the
strange quark mass.
The singular part of the free energy may be written as
\begin{equation}
\frac{f_s(T,\mu_q,h)}{T^4} =
A h^{1+1/\delta} f_f(z) \ ,\ z \equiv t/h^{1/\beta\delta} \; ,
\label{singular}
\end{equation}
where $\beta$ and $\delta$ are critical exponents of the 3-dimensional
$O(4)$ spin model~\cite{Engels} and
\begin{eqnarray}
t &\equiv& \frac{1}{t_0}\left( \frac{T-T_c}{T_c} +
\kappa_q \left( \frac{\mu_q}{T}\right)^2
\right)
\ ,
\nonumber \\
h &\equiv& \frac{1}{h_0} \frac{m_q}{T_c} \ .
\label{scalingfields}
\end{eqnarray}
Here $T_c$ is the critical temperature in the chiral limit
and $t_0$, $h_0$ are non-universal scale parameters (as is $T_c$).
We use the chiral transition temperature $T_c$ also to set the
scale for the explicit symmetry breaking, introduced by the non-vanishing
light quark masses\footnote{In Ref.~\cite{Ejiri} the strange
quark mass was used to set the scale for the symmetry breaking term.}.
The amplitude $A$ is fixed by the relation of the scaling function
for the free energy density, $f_f(z)$, to the more
commonly used scaling function $f_G(z)$, which characterizes the
scaling properties of the chiral condensate, or in general the order
parameter ($M$) in $O(4)$ symmetric models, $M = h^{1/\delta} f_G(z)$,
where
\begin{equation}
f_G(z) = -\left( 1+\frac{1}{\delta} \right) f_f(z) + \frac{z}{\beta\delta}
f'_f(z)  \; .
\label{fG}
\end{equation}
The scaling function $f_f(z)$ and its derivatives $f_f^{(n)}(z)$ have recently been
determined for $n\le 3$ using high precision Monte
Carlo simulations of the 3-dimensional $O(4)$ spin model and the known asymptotic
series expansions \cite{Engels_2011}.
We will use these results as a starting point
for a discussion of the generic structure of higher order cumulants of
the net baryon number fluctuations.

Note that the reduced temperature $t$, introduced in 
Eq.~(\ref{scalingfields}), depends explicitly
on the quark chemical potential. The constant
$\kappa_q\simeq 0.06$, which controls the curvature of the chiral phase 
boundary for small values of $\mu_q/T$, was recently determined in
a scaling analysis of (2+1)-flavor QCD \cite{Mukherjee}. A comparison
with other lattice results for $\kappa_q$~\cite{Endrodi,Philipsen}, suggests that
this parameter is only weakly dependent on the quark mass
and the number of flavors.

In this paper we focus on the properties of the net baryon number
fluctuations. The corresponding cumulants are obtained from Eq.~(\ref{freeenergy}) by taking
derivatives with respect to $\hmu_q = \mu_q/T$,

\begin{equation}
\chi_{n}^{B} = - \frac{1}{3^{n}}
\frac{\partial^{n}f/T^4}{\partial\hmu_q^n} \; .
\label{obs}
\end{equation}
From Eq.~(\ref{singular})
it is apparent that, in the vicinity of the critical temperature, the susceptibilities $\chi_{n}^{B}$ show a strong dependence on the explicit symmetry breaking term, the quark mass,
\begin{equation}
\chi_n^B \sim
\begin{cases}
-(2\kappa_q)^{n/2} h^{(2-\alpha -n/2)/\beta\delta} f_f^{(n/2)}(z)
& ,\ {\rm for}\ \mu_q /T = 0,\
{\rm and}\ n\ {\rm even} \\
- (2\kappa_q)^n \left( \frac{\mu_q}{T} \right)^n
h^{(2-\alpha -n)/\beta\delta} f_f^{(n)}(z)
&,\ {\rm for}\ \mu_q/T > 0\,
\end{cases}
\label{fluct_mass}
\end{equation}
where we used the scaling relation $2-\alpha = \beta\delta (1+1/\delta)$.
Because $\alpha$ is negative in the 3-dimensional  $O(4)$ universality class
($\alpha = -0.2131 (34)$ \cite{Engels}), the specific heat, or equivalently
the fourth order cumulants of the net baryon number fluctuations, 
does not diverge at the chiral transition
temperature, {\it i.e.} at $z=0$, in the chiral limit. At $\mu_q/T=0$ the first divergent cumulant 
is obtained for $n= 6$, while at $\mu_q/T > 0$ this happens already for $n= 3$.
We note that the singular structure appearing
in $n$-th order cumulants  for $\mu_q/T > 0$ is identical to that of $(2n)$-th order
cumulants at $\mu_q/T =0$, since
the chemical potential enters quadratically in the reduced
temperature $t$.
This characteristic has been used in Refs.~\cite{Asakawa, Skokov:2010}
to exploit properties of third  order cumulants at non-zero baryon
chemical potential as signatures for critical behavior.
In this case, however, the singular contributions
are suppressed by a factor $(\mu_q/T)^{n/2}$
relative to the sixth  order cumulants at $\mu_q/T =0$. Consequently, in the mean-field analysis of Ref.~\cite{Asakawa}, it was found that qualitative changes of the third-order cumulant, e.g. a change
of sign in $\chi_3^B$, is found only at rather large values of the chemical potential, $\mu_B/T > 4$.

Using Eq.~(\ref{fluct_mass}), one finds the leading
singularity in the chiral limit,
\begin{equation}
\chi_n^B \sim
\begin{cases}
-(2\kappa_q)^{n/2} |t|^{2-\alpha -n/2} f^{(n/2)}_\pm
& ,\ {\rm for}\ \mu_q /T = 0,\
{\rm and}\ n\ {\rm even} \\
-(2\kappa_q)^n \left( \frac{\mu_q}{T} \right)^n
|t|^{2-\alpha -n} f^{(n)}_\pm &,\ {\rm for}\ \mu_q/T > 0\; ,
\end{cases}\
\label{fluct}
\end{equation}
where
\begin{equation}
f^{(n)}_\pm = \lim_{z\rightarrow \pm \infty} |z|^{-(2-\alpha -n)} f_f^{(n)}(z)
\; .
\label{limit}
\end{equation}

\begin{figure}
\begin{center}
\vspace*{-0.8cm}
\includegraphics*[width=6.7cm]{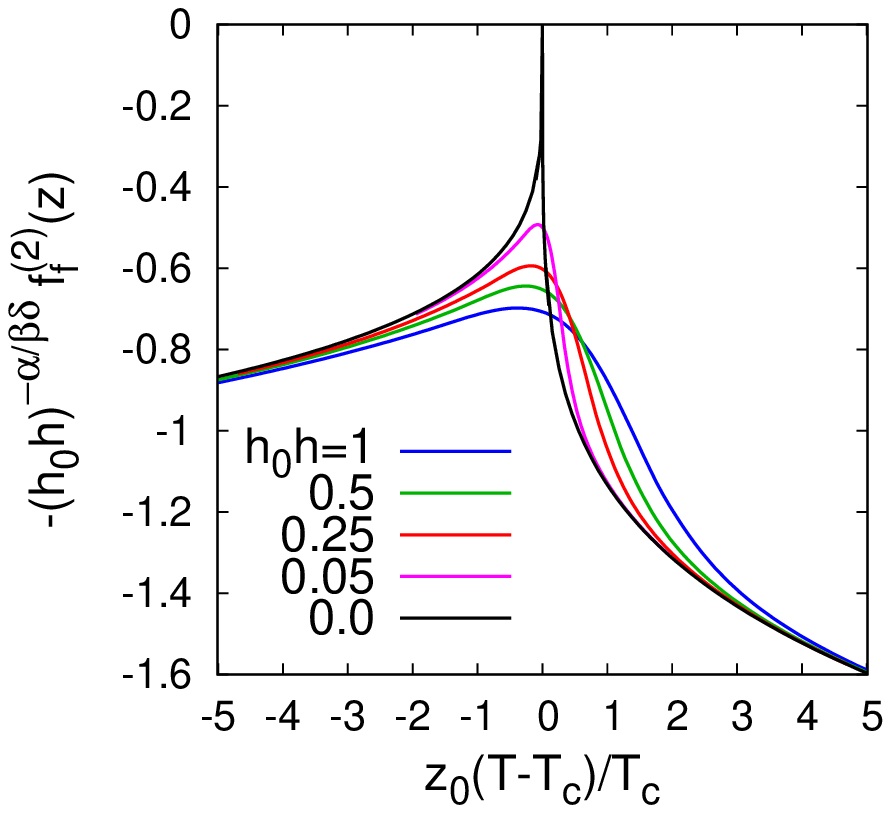}
\includegraphics*[width=6.7cm]{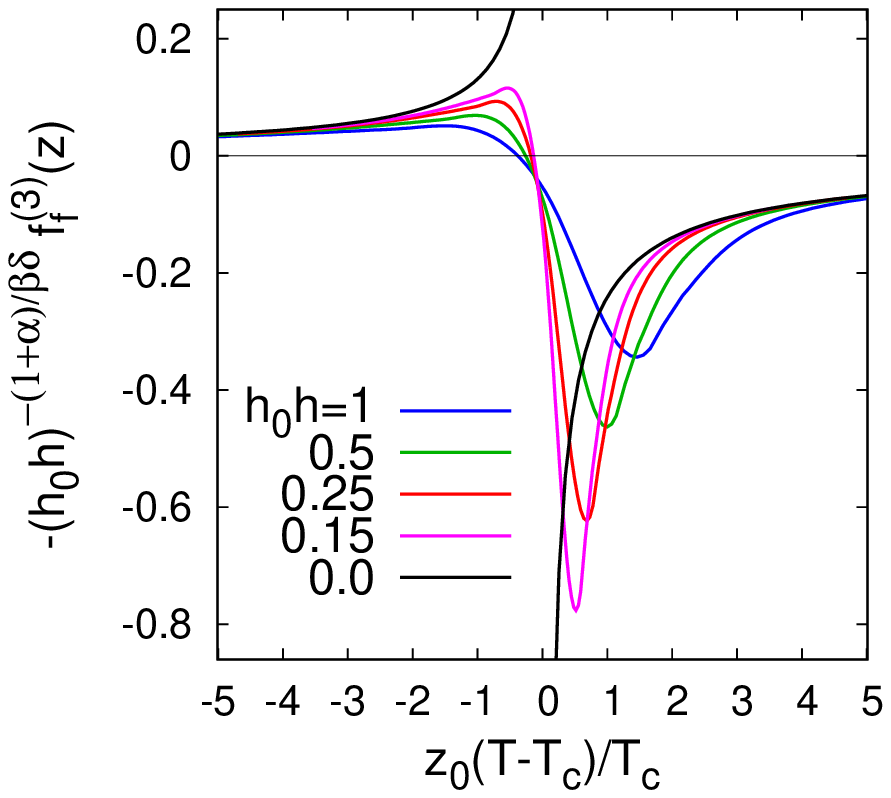}
\caption{
Scaling of the non-analytic contributions to
$\chi_4^B$ (left) and $\chi_6^B$ (right) arising from
second and third derivatives of the singular part of the free energy.
Shown are results for different values of the symmetry breaking
parameter $h_0h = m_q/T_c$; $h_0$ and $z_0=h_0^{1/\beta\delta}/t_0$ 
are non-universal scale parameters. Note that for $h_0h=1$ the
abscissa is the scaling variable $z$.
The corresponding curve thus directly shows the $O(4)$
scaling function. 
}
\label{fig:generic}
\end{center}
\end{figure}

\begin{figure}

\begin{center}
\includegraphics*[width=7.5cm]{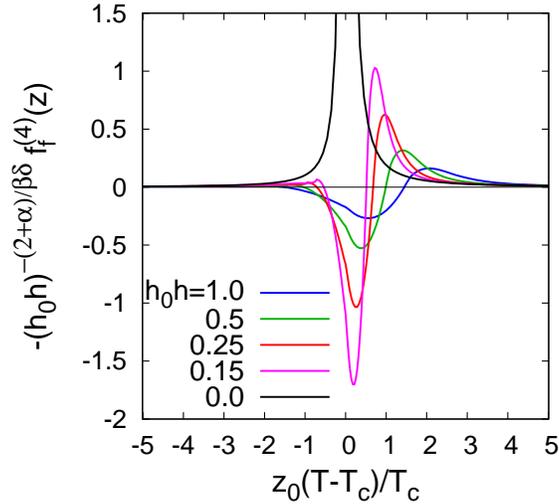}
\caption{Same as Fig.~\protect\ref{fig:generic} but for the
non-analytic contributions to $\chi_8^B$.
}
\label{fig:c8}
\end{center}
\end{figure}

The singular part of $\chi_4^B$, which is proportional to
$f_f^{(2)}$, has
the same singular structure as the specific heat; it is proportional to
the second derivative of the free energy with respect to temperature.
Thus, using the convention of Ref.~\cite{Engels_Cv}, we may write $\chi_4^B(t)$
in the chiral limit at $\mu_q/T=0$,
\begin{equation}
\chi_4^B(t) \sim \chi_{r} +\frac{A^\pm}{\alpha} |t|^{-\alpha} \; ,
\label{chi4}
\end{equation}
where $A^{+}$ is the amplitude above and $A^{-}$ below the critical temperature. The amplitudes $A^\pm$ are positive and the ratio $A^+/A^- \simeq 1.8$. This implies that
the cumulants $\chi_n^B(t)$  are positive for all $n > 4$ and $t<0$,
while for $t>0$ they alternate in sign. At non-zero values of the quark
mass, $h>0$, we thus expect $\chi_6^B$ to change sign in the transition
region and $\chi_8^B$ to do so twice. 
For a given $h>0$, this is reflected 
in the $z$-dependence of the scaling functions $f_f^{(n)}(z)$, as shown\footnote{
From the result for $f_f^{(3)}(z)$ \cite{Engels_2011} 
we also constructed an estimate
for the next derivative, $f_f^{(4)}(z)$, which required some smoothening
of the interpolations that entered the determination of $f_f^{(3)}(z)$.
The resulting scaling function and the resulting quark mass dependence
of $\chi_8^B(t)$ is shown in Fig.~\ref{fig:c8}.
We want to use it here to point out the qualitative structure that
does arise within the $O(4)$ universality class, but do not consider
this figure as being correct on the quantitative level, {\it i.e.} as far
as the accurate location of the minima and the height of peaks is concerned.}
in Figs.~\ref{fig:generic} and \ref{fig:c8}.
In fact, the temperature and quark mass dependence of the singular parts of
the net baryon number fluctuations is directly related to the scaling functions $f_f^{(n)}(z)$ of the 3-dimensional $O(4)$ model \cite{Engels_2011} .
Thus, the generic structure of the fourth and sixth order cumulants can
be obtained from the known $O(4)$ scaling functions, 
in the chiral limit as well as for non-zero values of the quark mass. 

In the chiral limit, the non-analytic contribution to $\chi_4^B$ vanishes 
at the chiral transition temperature, $t=0$. Consequently, in the transition region,
the regular terms dominate in $\chi_4^B$ .
Nonetheless, the non-analytic term in $\chi_4^B$ varies rapidly with
temperature, leading to a pronounced maximum in the transition region,
observed in lattice as well as model calculations.

The temperature at which $\chi_6^B$ changes sign is non-universal since it depends on the 
magnitude of the regular terms. However, in the scaling regime the location of the extrema and
the corresponding amplitudes follow universal scaling laws. Moreover, we note
that the positions of the two extrema of $f_f^{(3)}(z)$, $z^-<0$ and $z^+>0$, provide bounds on the 
critical temperature in the chiral limit.

It is evident from Figs.~\ref{fig:generic} and \ref{fig:c8} that the $O(4)$
scaling functions
show much more structure and a stronger quark mass dependence in the
symmetric phase, $t>0$, than in the broken phase, $t<0$. In the latter
case, the divergence at $t=0$ builds up much more slowly than on the high
temperature side. The scaling function $f_f^{(3)}(z)$ changes sign 
close to $z=0$, while $f_f^{(4)}(z)$ does so already
at $z^-< 0$. Furthermore, we note that the position of the maximum
of $f_f^{(3)}(z)$ is at $z^+\simeq 1.45$ \cite{Engels_2011}, which is close to
the peak position of the chiral susceptibility, $z_p\simeq 1.33(5)$
\cite{Engels}.
The location of these extrema define pseudo-critical temperatures $T_{6\pm}$,
which converge to that of the second order chiral phase transition
temperature $T_c$ in the chiral limit,
\begin{equation}
\frac{T_{6\pm}(m_q)}{T_c} = 1+ \frac{z^\pm}{z_0} \left( \frac{m_q}{T_c}\right)^{1/\beta\delta} \; ,
\label{T6}
\end{equation}
where $1/\beta\delta \simeq 0.55$. The proportionality constant $z_0$ is
uniquely determined in the scaling regime of QCD with two light quarks
\cite{Ejiri}
and is the same constant, which also controls the scaling of the
pseudo-critical temperature $T_\chi$ determined from the position of 
the peak in the chiral susceptibility,
\begin{equation}
\frac{T_{\chi}(m_q)}{T_c} = 1+ \frac{z_p}{z_0} \left( \frac{m_q}{T_c}\right)^{1/\beta\delta}
\; .
\label{Tchiral}
\end{equation}
With decreasing quark mass the minimum of $\chi_6^B$ decreases as,
\begin{equation}
\chi^B_{6,min} \sim -\ \left( \frac{m_q}{T_c}\right)^{-(1+\alpha)/\beta\delta}
\simeq -\ \left( \frac{m_q}{T_c}\right)^{-0.66} \; .
\label{c6_min}
\end{equation}
We also note that corrections to $\chi^B_{6,min}$ at non-zero
$\mu_q/T$ start at ${\cal O}( (\mu_q/T)^4)$. We thus expect that the
basic structure of higher cumulants persists also for $\mu_B/T>0$.

\subsection{Sixth order cumulant and the QCD transition temperature}

Based on the generic structure of the $O(4)$ scaling functions we thus can
understand the basic features of the temperature dependence of higher
cumulants of the baryon number fluctuations at vanishing baryon chemical
potential. We focus on the properties of the sixth order cumulant,
$\chi^B_{6}(T)$, or correspondingly the ratio of cumulants
$R_{6,2}^B(T) = \chi^B_{6}(T)/\chi^B_{2}(T)$. In the hadronic phase,
$\chi^B_{6}(T)$ first grows with increasing temperature. It exhibits a
maximum in the
hadronic phase, close to the transition region and then drops rapidly. In the entire
high temperature regime, $\chi^B_{6}(T)$ and consequently $R_{6,2}^B(T)$ remain
negative. Lattice calculations of $\chi^B_{6}(T)$ \cite{Allton,Schmidt}
suggest that in QCD with physical quark masses, these 
basic features, which are due to the singular terms in
$\chi^B_n$, persist. In particular,
$\chi^B_{6}(T)< 0$ in the vicinity of the pseudo-critical temperature
for chiral symmetry restoration.

We thus expect the sixth order
cumulant to be a very sensitive probe for the temperature at
which the freeze-out of hadrons in heavy ion collisions occurs.
In fact, it is conceivable
that hadronic freeze-out occurs in a temperature regime above 
the QCD phase transition temperature in the chiral limit, {\it i.e.} it may 
occur close to the pseudo-critical temperature that signals the onset of chiral
symmetry restoration. In that case we would expect to observe
\begin{equation}
R_{6,2}^B(T)\equiv \frac{\chi^B_{6}(T)}{\chi^B_{2}(T)} < 0
\; \text{at freeze-out at LHC and RHIC high energy runs.}
\nonumber
\label{prediction}
\end{equation}
This would be in striking contrast to conventional HRG model calculations,
which predict $R_{6,2}^B = 1$. A similar conclusion can be
drawn for the eighth order cumulant $\chi^B_8$ or, equivalently, the ratio
$R_{8,2}^B$.

\begin{figure}
\vspace*{-0.8cm}
\begin{center}
\includegraphics*[width=6.5cm]{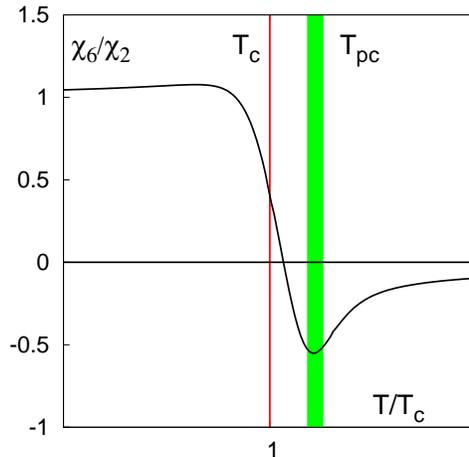}
\vspace*{-0.6cm}
\caption{Schematic plot of the temperature dependence of the ratio
of the sixth and second order cumulants of the net baryon number fluctuations
in units of the phase transition temperature $T_c$ in the chiral limit.
The vertical lines show the chiral phase transition temperature and
the pseudo-critical temperature $T_{pc}$, corresponding to a peak in
the chiral susceptibility in QCD with physical light quark
masses.
}
\label{fig:schematic}
\end{center}
\end{figure}

In more general terms we suggest that a determination of $R_{6,2}^B$
at large collision energies at RHIC or LHC will provide a characteristic signature for
the location of the freeze-out temperature relative to the QCD phase
transition. This is illustrated in Fig.~\ref{fig:schematic}, where we show a schematic plot of $R_{6,2}^B$ along with the critical temperature in the chiral limit ($T_c$) and the crossover
temperature for chiral symmetry restoration for physical
light quark masses ($T_{pc}$). 

In the following we extend the above considerations to the case of
non-vanishing baryon chemical potential. We do this in the framework
of a chiral model, analyzing the critical behavior of the
Polyakov loop extended quark meson (PQM) model in the functional
renormalization group approach \cite{Skokov:2010}. This allows us
to compute the dependence of higher cumulants on $\mu_q/T$ and $T$. We
determine the line of minima of $R_{6,2}^B$
in the $\mu_q$-$T$ plane and compare this with the
chiral transition line as well as with the pseudo-critical line obtained
at a physical pion mass.

\section{Higher cumulants of charge fluctuations at \boldmath $\mu_B/T > 0$}

The universal features of the 3-dimensional $O(4)$ scaling functions
discussed in the previous section also are reflected in effective chiral
model calculations where the Polyakov loop is coupled to
the fermion sector, thus generating many of the characteristics of the 
confinement-deconfinement transition \cite{Fukushima,Fukushima:strong,Ratti,Sasaki:2006ww}. 
The Polyakov loop extended quark meson (PQM) model~\cite{Schaefer:PQM} is one variant. It shares with QCD a global $O(4)$ symmetry, which
is spontaneously broken at low and restored at high temperatures.
Thus, in the light quark mass limit, this model reproduces the
universal scaling functions, discussed in the previous section.
We use the PQM model to implement the basic features
of the transition from hadronic matter to a quark gluon plasma into the
temperature  dependence of the cumulants of net baryon number fluctuations and
to analyze their properties also at $\mu_q/T > 0$.

The Lagrangian of the  PQM model   \cite{Schaefer:PQM},
\begin{eqnarray}\label{eq:pqm_lagrangian}
  {\cal L} &=& \bar{q} \, \left[i\gamma_\mu D^\mu - g (\sigma + i \gamma_5
  \vec \tau \vec \pi )\right]\,q
  +\frac 1 2 (\partial_\mu \sigma)^2+ \frac{ 1}{2}
  (\partial_\mu \vec \pi)^2
  \nonumber \\
  && \qquad - U(\sigma, \vec \pi )  -{\cal U}(\ell,\ell^{*})\ ,
\end{eqnarray}
involves interactions  of mesons  and gluons  with fermionic fields.
The coupling between the effective
gluon field and  quarks is implemented through the covariant derivative
\begin{equation}
 D_{\mu}=\del_{\mu}-iA_{\mu},
\end{equation}
where $A_\mu=g_s\,A_\mu^a\,\lambda^a/2$ and the spatial components of the gluon
field are neglected, i.e. $A_{\mu}=\delta_{\mu0}A_0$.

The purely mesonic potential of the model,
\begin{equation}
U(\sigma,\vec{\pi})=\frac{\lambda}{4}\left(\sigma^2+\vec{\pi}
^2-v^2\right)^2-c\sigma,
\end{equation}
contains a term linear in the $\sigma$ field, which  explicitly breaks the chiral symmetry and is used to reproduce the physical pion mass.

The  effective potential for the gluon field
is expressed in terms of the  thermal expectation values of the color trace of the  Polyakov loop and its conjugate,
\begin{equation}\label{pot}
 \frac{{\cal U}(\ell,\ell^{*})}{T^4}=
-\frac{b_2(T)}{2}\ell^{*}\ell
-\frac{b_3}{6}(\ell^3 + \ell^{*3})
+\frac{b_4}{4}(\ell^{*}\ell)^2\,
\end{equation}
where
\begin{equation}
\ell=\frac{1}{N_c}\vev{\Tr_c\, L(\vec{x})},\quad \ell^{*}=\frac{1}{N_c}\vev{\Tr_c\,
L^{\dagger}(\vec{x})},
\end{equation}
and
\begin{eqnarray}
   L(\vec x)={\mathcal P} \exp \left[ i \int_0^\beta d\tau A_4(\vec x , \tau)
  \right]\,.
\label{pot_l}
\end{eqnarray}
Here ${\mathcal P}$ stands for the path ordering, $\beta=1/T$ and $A_4=i\,A_0$.
The potential ${\cal U}(\ell,\ell^{*})$ preserves the $Z(3)$ symmetry of the gluonic sector of QCD.

In the Appendix, we give further details on the choice of parameters for the
Polyakov loop potential and present the relevant calculational steps within
the functional renormalization group (FRG) approach
\cite{Wetterich, Morris,Ellwanger,Berges:review}. Within this formalism we compute
the free energy density of the PQM model
\cite{Skokov:2010,Skokov:2010wb},
\begin{equation}
f_{\rm PQM}(\ell, \ell^*;T, \mu) \equiv  \Omega(\ell, \ell^*;T, \mu)
= -\frac{T}{V}\ln Z(\ell, \ell^*;T, \mu)\; ,
\label{fPQM}
\end{equation}
as a function of temperature,
chemical potential as well as the Polyakov loop variables, $\ell$ and $\ell^*$.
The latter two are then fixed by the stationarity condition:
\begin{eqnarray}
\label{eom_for_PL_l}
\frac{ \partial   }{\partial \ell} \Omega(\ell, \ell^*;T, \mu)  =0,~~
\frac{ \partial   }{\partial \ell^*}  \Omega(\ell, \ell^*;T, \mu)   =0 \; .
\label{eom_for_PL_ls}
\end{eqnarray}
The cumulants of the net baryon number fluctuations are obtained
by taking suitable derivatives of the free energy, as specified in Eq.~(\ref{obs}). In practice
these derivatives have been implemented directly into the analysis
of the flow equations (see Appendix).
\begin{figure}
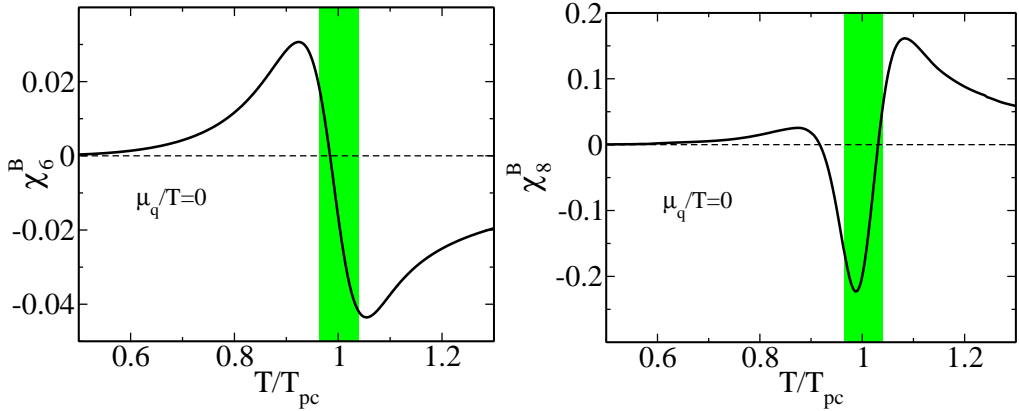

\begin{center}
\includegraphics*[width=6.5cm]{c_6.eps}
\includegraphics*[width=6.8cm]{c_8.eps}
\end{center}
\caption{The sixth and eighth order cumulants of the
net baryon number  fluctuations at $\mu_q/T=0$ in the PQM
model. The temperature is given in units of the pseudo-critical
temperature $T_{pc}(m_\pi)$  corresponding to a maximum of the the chiral susceptibility.
The shaded area indicates the chiral crossover region.
}\protect\label{fig:PQM}
\end{figure}

\begin{figure}[t]
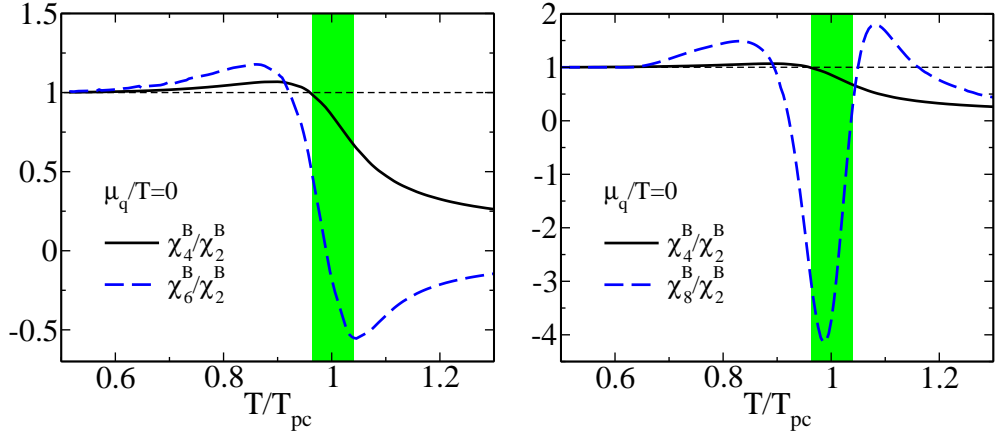

\begin{center}
\vspace*{-0.5cm}
\includegraphics*[width=6.5cm,]{c6c2.eps}\hspace*{0.3cm}
\includegraphics*[width=6.2cm,]{c8c2.eps}
\vskip 0.0cm\caption{The temperature  
dependence of the  fourth, sixth and eighth order cumulants of the net baryon 
number fluctuations $\chi_{n}^{B}$ relative to the second order one. The 
temperature is given in units of the chiral crossover temperature. The shaded 
area indicates the region of the chiral crossover transition at $\mu_q/T=0$.  
The calculations were done in the PQM model within the FRG approach.
}
\label{fig:PD0}
\end{center}
\end{figure}
\begin{figure}[t]
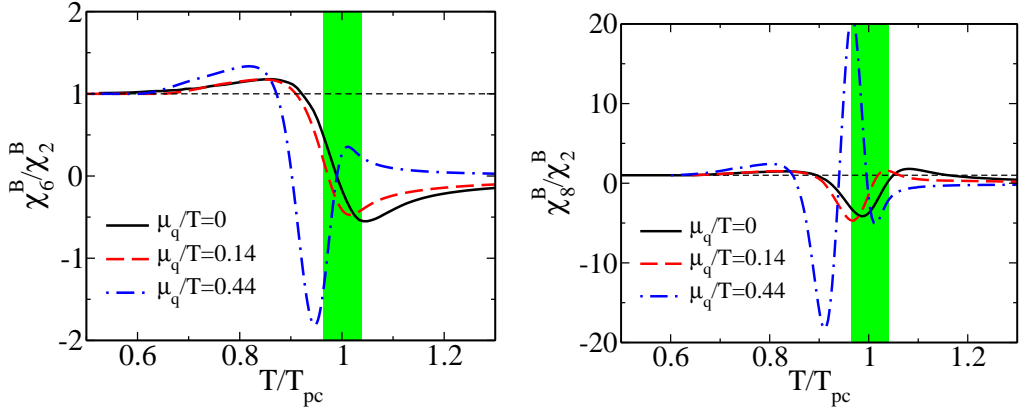

\begin{center}
\hspace*{-0.5cm}
\includegraphics*[width=6.5cm,]{c6c2mu02.eps}\hspace*{0.3cm}
\includegraphics*[width=6.5cm,]{c8c2mu02full.eps}
\vskip 0.0cm\caption
{
Comparison  of the temperature dependence of the ratios $\chi_{6}^B$/$\chi_{2}^B$ and $\chi_{8}
^B/\chi_{2}^B$ for various $\mu_q/T$ corresponding to values at chemical
freeze-out in heavy ion collisions at RHIC.
The shaded
area indicates the region of the chiral crossover transition at $\mu_q/T=0$.
}
\label{fig:PD}
\end{center}
\end{figure}

In Fig.~\ref{fig:PQM} we show the sixth and eighth order cumulants of
the net baryon number fluctuations computed at $\mu_q/T=0$ within the PQM
model for physical values of the pion mass. The basic features dictated
by $O(4)$ symmetry restoration, as discussed in the previous sections,
are readily identified in the figure.
Moreover, the positions of the two extrema of $\chi_6^B$ correspond 
approximately to the zeros
of $\chi_8^B$. This confirms that in the transition region, two derivatives
with respect to $\mu_q/T$ are indeed equivalent to one derivative with
respect to $T$.

From these calculations, as well as from calculations of the lower order
cumulants $\chi^B_2$ and $\chi^B_4$, we obtain the ratios $R_{n,m}^B$ of
the $n$-th and $m$-th cumulants.
Results obtained for $\mu_q/T=0$ and $\mu_q/T > 0$ are shown
in Figs.~\ref{fig:PD0} and \ref{fig:PD}, respectively. We note that these
ratios approach unity at low temperatures, as it is the case also in the
hadron resonance gas model. In the transition region, they reflect the
expected  $O(4)$ scaling properties; they have a shallow maximum close
to the transition region before they drop sharply. In particular, they
show pronounced
minima with $R_{n,2}^B<0$ in the vicinity of the chiral crossover
temperature. The exact location of these minima and their depth is
to some extent model dependent. However,
we note, that in the transition region the second order cumulant
used  in these ratios for normalization is dominated by non-singular
contributions which are positive.
The minima
in $R_{n,2}^B$  therefore mainly reflect the strong temperature dependence of
higher cumulants $\chi_{n}^B$. We also note that these minima become more pronounced
with increasing $\mu_q/T$. In fact, the structure of e.g. $R_{6,2}^B$
becomes similar to that of $\chi_{8}^B$  at large $\mu_q/T$. This is easily understood
in terms of the Taylor expansion of $R_{6,2}^B$, where 
the dominant correction at non-zero $\mu_q/T$ is due to $\chi_{8}^B$ ,
\begin{eqnarray}
R_{6,2}^B(\mu_q/T) &=& R_{6,2}^B(0) +\frac{1}{2}\left( \frac{\mu_q}{T}
\right)^2 \left( R_{8,2}^B(0) -R_{6,2}^B(0) R_{4,2}^B(0) \right)
\nonumber \\
&&+ {\cal O}((\mu_q/T)^4) \; .
\label{Taylor}
\end{eqnarray}
This
also makes it clear why for $\mu_q/T >0$ the location of the minimum
of $R_{6,2}^B(\mu_q/T)$ is shifted to lower temperatures relative
to that of the chiral crossover temperature. Similarly, at non-zero $\mu_q/T$, 
the ratio $R_{8,2}^B(\mu_q/T)$ shows more pronounced oscillations in the
transition region, due to contributions from higher order cumulants, which 
oscillate more rapidly in the transition region. The amplitude of the maximum 
at high temperatures becomes compatible in magnitude with that of the minimum.

\section{Freeze-out  and the QCD transition}

The analysis of universal scaling functions that control the thermodynamics
in the vicinity of a phase transition in the universality class of
3-dimensional $O(4)$ symmetric theories (section 2), 
of model (section 3) 
and of lattice calculations
\cite{Schmidt,Cheng} suggest that at vanishing baryon chemical potential
the sixth order cumulant of the net baryon number fluctuations is negative
in the entire high temperature phase. In fact, the $O(4)$ scaling functions
for higher cumulants turn negative already in the vicinity of the chiral
($m_q=0$) critical temperature, {\it i.e.} below the crossover
temperature  ($m_q>0$), which is relevant for the transition
at non-zero values of the light quark masses. The regular terms in the QCD 
free energy may shift the onset of the negative regime to higher temperatures. 
However, model \cite{Schaefer:2009st}
as well as lattice \cite{Schmidt,Cheng} calculations suggest that
the regime of negative sixth order cumulants starts below but close to the
QCD crossover transition temperature.

\begin{figure}[t]
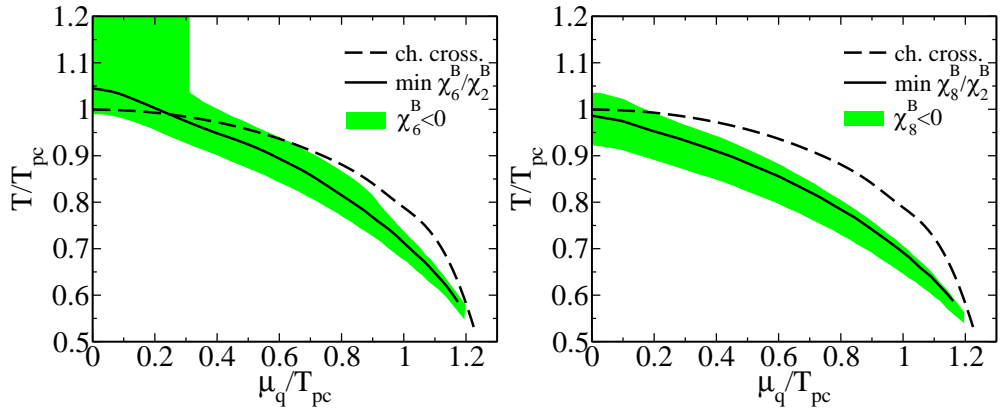

\begin{center}
\includegraphics*[width=6.5cm,]{negative_c6.eps}
\includegraphics*[width=6.5cm,]{negative_c8.eps}
\vskip -0.0cm\caption
{
The chiral crossover line  [dashed line] and the first minima
in $\chi_6^B$ (left) and $\chi_8^B$ (right) [solid line]. The bands show
the parameter range for which $\chi_6^B$  and $\chi_8^B$, respectively,
are negative in the neighborhood of these minima. 
}
\label{fig:lines}
\end{center}
\end{figure}

At non-zero baryon chemical potential, the temperature interval of negative
$\chi_6^B$, or equivalently $R^B_{6,2}(\mu_q/T)$,  shrinks and
follows the crossover transition line. This is illustrated in
Fig.~\ref{fig:lines},
which shows  the temperature interval, closest to the hadronic phase,
where the sixth
and eighth order cumulants of the net  baryon number fluctuations are negative, 
as obtained in the FRG approach to the PQM model.
It is evident that the sixth order cumulant $\chi_6^B$ is
negative in a wide range of temperatures which
extends into the symmetry broken phase. This is even more the case for
the eighth order cumulant as expected from the structure of the corresponding
$O(4)$ scaling function. Except for a small range of chemical potential
values close to $\mu_q/T = 0$, the eighth order cumulant is, however,
positive again on the crossover line.

\section{Discussion and Conclusions}

We have shown that higher order cumulants of the net baryon number fluctuations
are sensitive probes for the analysis of freeze-out conditions in heavy ion
collisions and may allow to clarify their relation to the QCD phase transition.
This is the case at LHC energies as well as at the entire regime
of beam energies covered by the low energy run at RHIC. If in heavy ion
collisions, particles are produced from a thermalized system, the analysis
of higher cumulants of the net baryon number fluctuations does provide constraints on the
location of the freeze-out
temperature relative to the chiral transition temperature. We find that the
most robust statements, which become rigorous in the chiral limit,
can be made for the sixth order cumulants of charge fluctuations for small values of the quark chemical potential:

{\it If freeze-out occurs close to the chiral crossover temperature the
sixth order cumulant of the net baryon number fluctuations will be negative
at LHC energies as well as for RHIC beam energies $\sqrt{s_{NN}}\ \gsim\ 60$~GeV,
corresponding to $\mu_B/T\lsim 0.5$. This is in contrast to hadron
resonance gas model calculations which yield a positive sixth order
cumulant.}

We also note that the basic features discussed here for
net baryon number fluctuations also carry over one-to-one to 
electric charge fluctuations. In the chiral limit, the most singular
component in the cumulants of electric charge fluctuations is proportional
to the singular part of the corresponding cumulant of
net baryon number fluctuations. In fact, lattice~\cite{Cheng} and 
model~\cite{Fu:2009wy} calculations of the
sixth order cumulant of electric charge fluctuations at $\mu_q/T=0$
show that this cumulant is negative in the high temperature phase of
QCD. The ratio of the sixth and second order cumulants,
$\chi_6^Q/\chi_2^Q$ rapidly drops in the transition region from
the HRG value $(\chi_6^Q/\chi_2^Q)_{HRG}\simeq 10$ to about $-4$
at the chiral crossover temperature. Also this ratio is therefore a
sensitive probe of the conditions at freeze-out and
their relation to the critical behavior in strongly interacting matter.

We finally comment on the fourth order cumulants, in particular on the
ratio $\chi_4^B/\chi_2^B$ recently measured by STAR \cite{STAR} as well
as the corresponding ratio for cumulants for electric charge fluctuations,
$\chi_4^Q/\chi_2^Q$. The former is consistent
with the HRG value $(\chi_4^B/\chi_2^B)_{HRG}=1$ \cite{Karsch_2010,Gavai_2010}. 
If freeze-out
occurs in the hadronic phase, one expects to find also large values
for ratios of cumulants of electric charge fluctuations,
$(\chi_4^Q/\chi_2^Q)_{HRG}\simeq 1.8$ \cite{Karsch_2010}. On the other hand,
if freeze-out occurs at the crossover temperature, also these ratios
will deviate from the HRG values. Lattice calculations suggests that
both ratios drop rapidly in the transition region, but stay positive
also in the high temperature phase \cite{Cheng}.

\begin{table}[t]
\begin{center}
\begin{tabular}{|l|c|c|c|c|}
\hline
~freeze-out conditions & $\chi_4^B/\chi_2^B$ &$\chi_6^B/\chi_2^B$ &
$\chi_4^Q/\chi_2^Q$ &$\chi_6^Q/\chi_2^Q$ \\
\hline
 HRG&  1 & 1 & $\sim$ 2& $\sim$ 10 \\
QCD: $T^{freeze}/T_{pc}\lsim 0.9$ &  $\gsim 1$ & $\gsim 1$ & $\sim 2$&
$\sim 10$ \\
QCD: $T^{freeze}/T_{pc}\simeq 1$ & $\sim 0.5$ & $ < 0$ & $\sim 1 $ & $< 0$ \\
\hline
\end{tabular}
\end{center}
\caption{\label{tab:scenario}
Values for ratios of cumulants of net baryon number (B) and
electric charge (Q) fluctuations for the case that freeze-out appears
well in the hadronic phase (third row) or in the vicinity of the chiral
crossover temperature (fourth row). We give results
based on current lattice calculations \cite{Schmidt,Cheng} and on the calculations
presented here. In the second row we give results
of a HRG model calculation \cite{Cheng}. We also note that unlike the
cumulants of net baryon number fluctuations the ratios of cumulants of
electric charge fluctuations vary somewhat as a function
of the baryon chemical potential along the freeze-out line.
}
\end{table}

In Table~\ref{tab:scenario}  we summarize two different freeze-out scenarios
characterized by various ratios of the cumulants of the net baryon number and electric
charge fluctuations, respectively. These scenarios assume that in heavy ion
collisions the freeze-out of hadrons occurs from a thermalized system
characterized by the phenomenologically determined freeze-out curve
$T_f(\mu_B^f)$. Table~\ref{tab:scenario} shows, that ratios of cumulants of 
charge fluctuations  are very sensitive to possible
differences in freeze-out and crossover temperature.
Finally, we note that these results do not account for possible finite volume
effects  nor for possible effects of the evolution from chemical towards thermal freeze-out
in heavy ion collisions.

\section*{Acknowledgments}
We gratefuly acknowledge discussions with J\"urgen Engels on the $O(4)$
scaling functions.
The work of F.K. was supported in part by contract DE-AC02-98CH10886
with the U.S. Department of Energy, by the BMBF under grant 06BI401 and the
GSI Helmholtzzentrum f\"ur Schwerionenforschung under grant BILAER. K.R. acknowledges
partial support by the Polish Ministry of Science (MEN). B.F. and K.R. were supported in part by  the 
ExtreMe Matter Institute (EMMI). V.S. acknowledges support by the Frankfurt Institute for 
Advanced Studies (FIAS).

\addcontentsline{toc}{section}{Acknowledgements}

\appendix
\section*{Appendix}

Here we summarize the basic steps involved in the calculation of the free
energy of the PQM model within the functional renormalization group (FRG)
approach. We also describe the methods used to obtain the FRG results 
for the cumulants of the net baryon number fluctuations. Further details on these 
calculations are given in Refs.~\cite{Skokov:2010wb,Skokov:2010,stokic}.

In the PQM model Lagrangian introduced in
 Eqs.~(\ref{eq:pqm_lagrangian}-\ref{pot_l}), we need to specify the
parameters used in the Polyakov loop potential (\ref{pot}).
These parameters
were chosen  to reproduce the equation of state of pure $SU(3)$ lattice
gauge theory,
\begin{eqnarray}
\hspace{-4ex}
  b_2(T) &=& a_0  + a_1 \left(\frac{T_0}{T}\right) + a_2
  \left(\frac{T_0}{T}\right)^2 + a_3 \left(\frac{T_0}{T}\right)^3\,,
\end{eqnarray}
with  $a_0 = 6.75$, $a_1 = -1.95$, $a_2 = 2.625$, $a_3 = -7.44$, $b_3 = 0.75$,  $b_4 = 7.5$
and $T_0 = 270$ MeV.

The  FRG flow equation for the thermodynamic potential in general contains
mesons,  quarks  and Polyakov loops as  dynamical fields. However, in the
current  calculation, we treat the Polyakov loop as a background field which is
introduced self-consistently on the mean-field level.
Following previous work \cite{Skokov:2010,Skokov:2010wb}, we formulate the
flow equation for the scale-dependent grand canonical thermodynamic potential density  for
quarks and mesons 
\begin{eqnarray}
\nonumber
\del_k \Omega_k(\ell, \ell^*; T,\mu_q)&=&\frac{k^4}{12\pi^2}
 \left\{ \frac{3}{E_\pi} \Bigg[ 1+2n_B(E_\pi; T)  \Bigg]
 +\frac{1}{E_\sigma} \Bigg[ 1+2n_B (E_\sigma; T)
    \Bigg]   \right. \\  && \left. -\frac{4 N_c N_f}{E_q} \Bigg[ 1-
N(\ell,\ell^*;T,\mu_q)-\bar{N}(\ell,\ell^*;T,\mu_q)\Bigg] \right\}.
\label{eq:frg_flow}
\end{eqnarray}
Here
\begin{equation}
n_B(E_{\pi,\sigma};T)=\frac{1}{\exp({E_{\pi,\sigma}/T})-1},
\end{equation}
is the bosonic  distribution function, with the pion and sigma energies,
\begin{equation}
E_\pi = \sqrt{k^2+\overline{\Omega}^{\,\prime}_k}\;~,~ E_\sigma
=\sqrt{k^2+\overline{\Omega}^{\,\prime}_k+2\rho \,\overline{\Omega}^{\,
\prime\prime} _k},
\end{equation}
where $\overline{\Omega_k}=\Omega_k+c\sigma$ and  primes denote the  derivatives  with respect to $\rho=\sigma^2/2$.
The momentum  distributions of quarks $N$ and antiquarks $\bar{N}$ are modified owing to the coupling to gluons,
\begin{eqnarray}
N(\ell,\ell^*;T,\mu_q)&=&\frac{1+2n_{\bar q}\ell^*+ n_{\bar q}^2\ell}{1+3 n_{\bar q}^2\ell+
3n_{\bar q}\ell^*+n_{\bar q}^3},   \nonumber\\
\bar{N}(\ell,\ell^*;T,\mu_q)&=&N(\ell^*,\ell;T,-\mu_q),
\label{n2}
\end{eqnarray}
where $E_q =\sqrt{k^2+2g^2\rho }$ is the quark energy  and $n_{\bar q}(E_q;T,\mu_q)=\exp[\beta(E_q-\mu_q)]$.

We solve the flow equation for  $\Omega_k(T,\mu_q)$ numerically by
expanding it in
powers of field variable, $\rho$,  around the scale dependent  potential minimum $\rho_{k} = \sigma_{k}^2/2$,
\begin{equation}\label{eq:taylor}
\bar{\Omega}_k(T,\mu_q)=\sum_{m=0}^{M} \frac{a_{m,k}(T,\mu_q)}{m!}(\rho-\rho_{k})^m.
\end{equation}
The location of the scale dependent minimum of thermodynamic potential  $\Omega_k(T,\mu_q)$
is determined by the stationarity condition
\begin{equation}
\left. \frac{d \Omega_k }{d \sigma} \right|_{\sigma = \sigma_{k}} = \left. \frac{d \bar{\Omega}_k }{d \sigma} \right|_{\sigma = \sigma_{k}} -c =0 .
\label{minimum_of_th_p}
\end{equation}
Truncating the expansion in Eq.~(\ref{eq:taylor}) at $M=3$,
 Eq.~(\ref{eq:frg_flow}) yields the flow equation  for the Taylor  coefficients and field variable, $\rho_k$, as
\begin{eqnarray}
\partial_ka_{0,k}&=&\frac{c}{\sqrt{2\rho_{k}}}\,\del_k\rho_{k}+\del_k\Omega_k,\label{eq:trunc1}
\\
\del_k\rho_{k}&=&-\frac{1}{c/(2\rho_{k})^{3/2}+a_{2,k}}
\del_k\Omega'_k,\label{eq:trunc2}
\\
\del_ka_{2,k}&=&a_{3,k}\, \del_k\rho_{k}+\del_k\Omega''_k,
\\
\del_ka_{3,k}&=&\del_k\Omega'''_k.\label{eq:trunc4}
\end{eqnarray}
These differential equations are solved
numerically with the initial cutoff $\Lambda=1.2$ GeV (for details see 
Refs.~\cite{Skokov:2010,Skokov:2010wb}).
The initial conditions for the flow  are chosen
to  reproduce  the  pion mass in vacuum $m_{\pi}=138$ MeV, the pion decay constant
$f_{\pi}=93$ MeV, the sigma mass $m_{\sigma}=600$ MeV and the constituent quark mass $m_q=310$ MeV at the scale $k=0$.

By construction, the solution of Eq.~(\ref{eq:frg_flow}), $\Omega_k$,
computed at the minimum $\rho = \rho_k$ describes the scale-dependent
thermodynamic potential density, where the contribution of soft quark and meson  modes with $|\vec{q}|<k$ is suppressed \cite{Skokov:2010wb}.
Integrating the flow equation from $k=\Lambda$ to  $k\to0$, we obtain
the thermodynamic
potential which accounts for all momentum modes $|\vec{q}|<\Lambda$.
The solution of equations (\ref{eq:trunc1}-\ref{eq:trunc4}),  yields the thermodynamic potential density
for   quarks and mesons, $\Omega_{k\to0} (\ell, \ell^*;T, \mu_q)$,
as a function of the Polyakov loop variables $\ell$ and $\ell^*$. The  full thermodynamic potential density  $\Omega(\ell, \ell^*;T, \mu_q)$ in the PQM model, 
including quarks, mesons and
gluons,
is obtained by adding   the
effective gluon  potential to  $\Omega_{k\to0} (\ell, \ell^*;T, \mu_q)$,
\begin{equation}
\Omega(\ell, \ell^*;T, \mu_q) = \Omega_{k\to0} (\ell, \ell^*;T, \mu_q) + {\cal U}(\ell, \ell^*),
\label{omega_final}
\end{equation}
where  at a given temperature and chemical potential, the Polyakov loop variables are determined by the
stationarity conditions:
\begin{eqnarray}
\label{eom_for_PL_l}
\frac{ \partial   }{\partial \ell} \Omega(\ell, \ell^*;T, \mu_q)  =0,~~
\frac{ \partial   }{\partial \ell^*}  \Omega(\ell, \ell^*;T, \mu_q)   =0.
\label{eom_for_PL_ls}
\end{eqnarray}

The thermodynamic potential  (\ref{omega_final}) does not contain the contributions of  statistical modes with momenta
larger than the cutoff $\Lambda$.
In order to obtain the correct high-temperature behavior of
the thermodynamics we supplement the
FRG potential with the  contribution of  high-momentum states with $|\vec{q}|> \Lambda$.
This contribution to the
flow is approximated by that of massless quarks
interacting with the Polyakov loops \cite{Skokov:2010,Skokov:2010wb},
\begin{eqnarray}\label{eq:qcdflow}
\del_k \Omega_k^{\Lambda}(T,\mu_q)&=&-\frac{N_c N_f k^3}{3\pi^2}
\\
&& \hspace*{-1cm} \Big[ 1-
N(\ell,\ell^*;T,\mu_q)-\bar{N}(\ell,\ell^*;T,\mu_q)\Big].\nonumber
\end{eqnarray}

The complete thermodynamic potential of the PQM model  is then obtained by integrating
Eq.~(\ref{eq:qcdflow})  from $k=\infty$ to
$k=\Lambda$, where we switch to the PQM flow
equation (\ref{eq:frg_flow}). The thermodynamic potential thus obtained can be  used to
  explore the properties of different cumulants of the net baryon number fluctuations.

The flow equation yields the thermodynamic potential as a function of temperature
and chemical potential. Different cumulants  of the net baryon number fluctuations $\chi_n^B$, 
Eq. (\ref{obs}), can then  in principle be obtained by
explicit numerical differentiation.
However,  owing to the numerical errors accumulated in the solution of the  flow equations,
this method does not provide reliable results.
Therefore, we compute the  higher  cumulants by analytically  differentiating   the
flow equations~(\ref{eq:trunc1}-\ref{eq:trunc4})  and ~(\ref{eq:qcdflow})
with respect to $\hat\mu_q$. This 
 defines flow equations for $\chi_1^B$ and  $\chi_2^B$, which are solved using the solution of Eqs.~(\ref{eq:trunc1}-\ref{eq:trunc4}) as input. Subsequently, the higher cumulants of the net baryon number fluctuations up to eighth order are computed by 
numerical differentiations of $\chi_2^B$.

\end{document}